\documentstyle[12pt]{article}
\normalbaselines
\begin{document}
\bibliographystyle{unsrt}
\normalsize
\indent
\\
Valery P. Karassiov and Andrei B. Klimov\\
P.N. Lebedev Physical Institute\\
AN ALGEBRAIC APPROACH TO SOLVING EVOLUTION PROBLEMS IN SOME
NONLINEAR  QUANTUM MODELS (11pages)
\\
\begin{abstract}
 A new general Lie-algebraic approach is proposed to solving evolution tasks
 in some nonlinear problems of quantum physics with polynomially deformed Lie
 algebras $su_{pd}(2)$ as their dynamic symmetry algebras.
 The method makes use of an expansion of the evolution operators
 by power series in the $su_{pd}(2)$ shift operators and a (recursive)
reduction
 of finding coefficient functions to solving auxiliary exactly solvable $su(2)$
 problems with quadratic Hamiltonians.
\end{abstract}
\\
 \section{Introduction}
 As is well-known, nonlinear models (with more than quadratic Hamiltonians) are
 intensively examined now in different branches of modern quantum physics. At
the sa
me time,
 at present there are not adequate universal techniques for  analyzing and
solving t
hem
 unlike the case of models with quadratic Hamiltonians which are efficiently
solved
by using
 group-theoretical and Lie-algebraic methods (see, e.g., [1,2]). But recently a
new
class of
nonlinear or deformed Lie algebras has  been introduced in
different areas of modern physics (see, e.g., [3-8] and references therein).
All these objects may be considered as extensions $g_{d}  = h + v$ of usual Lie
 alg
ebras
$h = \{ E_{c}\}$ by their ireducible tensor operators $v = \{ V_{c}\}$
satisfying the commutation relations(CRs): $[ E_a, V_b]  = \sum_{c}\tau_{ab}^c
V_c,

[ V_a , V_b ]  =  f_{ab}(E_c),  V_a \in  v, E_c \in h $
where $\tau_{ab}^c$  are matrix elements of operators $V_{c}$ and $f_{ab}(E_c
)$ are

some power series in generators of the subalgebra $^{"}h^{"}$ only.

Specifically, in [8,9] it was shown that such deformed Lie
algebras arise in a natural manner in composite many-body physics
models with Hamiltonians $H$ having invariance groups $G_{inv}$
($[H, G_{inv}]= 0$) and presented (via a
$G_{inv}$-invariant generalization of the Jordan mapping[9]) as
linear forms in elements of finite sets $I(G_{inv})$ of basic
invariants of groups $G_{inv}$. The sets $I(G_{inv})$ with commutators
$[A,B]=AB-BA$ generate, in general, nonlinear Lie algebras $g_{pd} $ with the
above
structure functions $f_{ab}(E_c )$ being polynomials.
Algebras $g_{pd}$ retain certain properties of familiar Lie
algebras and form together with $G_{inv}$
generalized Weyl-Howe's dual pairs $(G_{inv}, g_{pd})$ [8,9] which
act complementarily on the Hilbert spaces $L(H)$ of quantum
states of models under study, providing decompositions
$ L(H)  = \sum_{[l_i]} \oplus L([l_i ]) $
of $L(H)$ into direct sums of the $(G_{inv}, g_{pd})$- invariant
subspaces $L([l_i ])$ which evolve indepedently in time when governing
by Hamiltonian $H$.

All that opens up some ways of applications of the $g_{pd}$
formalism to solving physical tasks by analogy with usual Lie algebras[1,2].
However, there exist some peculiarities of applications of nonlinear Lie
algebras in comparison with those of familiar Lie algebras.
Specifically, for algebras $g_{pd}$ there do
not exist satisfactory definitions of the Wigner $D$-function analogs via
matrix ele
ments
of $g_{pd}$ exponentials (which are non-analytical) or of the group orbit type
gener
alized coherent states[8].
Therefore, for solving both spectral and evolution tasks we can use only direct
algebraic methods. Since spectral problems were examined in detail within this
approach in the papers [8,9] (see also [10] where a similar approach was used,
but w
ithout
exploiting deformed algebras, for solving the quantum Ablowitz-Ladik system),
below

the main attention is focused on applications to solving
temporal evolution problems.
%In Section 2  we  recapitulate  key points of the approach[8,9]
%by analyzing some (widespread in applications) quantum models with essentially
%%nonl
inear
%Hamiltonians, whose dynamic symmetry algebras $ g^{ds}$ are nonlinear Lie
%%algebras

%$sl_{pd}(2)$, give a "polar" decomposition of the  $sl_{pd}(2)$ shift
%%operators $V_
{\pm}$
%and write down a representation of temporal evolution operators $U_H(t)$ by
%%power s
eries in $V_{\pm}$
%and equations for finding appropriate coefficient functions. In  Section 3 we
%%solv
e these
%equations and obtain some new analytical solutions for $U_H(t)$ of the models
%%under
 study.
%In conclusion we briefly discuss some possible ways of further developments.
\section{Polynomial deformations $sl_{pd}(2)$ of the $sl(2)$ Lie algebra
     in some nonlinear problems of quantum physics}
%Evoluton operator and its expansion in power series (Ansatz by VPK);
%Difference equations for coefficient fuctions of this expansion.

    Let us consider (following to [8,9]) quantum models with Hamiltonians of
the for
m
 %=====
$$ a) H_1 = \sum _{i=1}^2 \omega_i a_i^+ a_i  +  g_1 (a^+_1)^n (a_2)^m +  g_1^*
(a_1
)^n (a^+_2)^m, 0 \leq m \leq n,
\eqno (2.1a)$$
$$ b) H_2 = \sum _{i=1}^n \omega_i a_i^+ a_i  + \omega_0 a_0^+ a_0  + g_2
(a^+_1...a
^+_n) (a_0)^m +  g_2^* (a_1...a_n) (a^+_0)^m,
\eqno (2.1b)$$
$$c) H_3 = \omega_1 a_1^+ a_1  + \sum _{i=1}^N  [\sigma_0 (i) \epsilon + g_3
\sigma_
{+} (i)(a _1)^n + g_3^* \sigma_{-} (i)(a _1^+)^n ]
 \eqno (2.1c)$$
%=============
where $g_i$ are some constants or time-dependent functions, $a_{i}, a _{i}^+ $
are boson operators, $\sigma_{\alpha}(i)$ are Pauli matrices, and
non-quadratic parts of $H_i$  describe different multiphoton processes
of scattering and frequency conversion
as well as n-photon point-like Dicke models of the matter-radiation
interactions [8]
{}.

     By a direct inspection of eqs (2.1) one can find that all these
Hamiltonian pos
sess
the symmetry groups $G_{i}= G_{inv}(H_i): G_1 = C_n \bigotimes C_m \bigotimes
U_1(1)
,
 G_2 = C_n \bigotimes U_2^0(1)\bigotimes U_2^1(1)...\bigotimes U_2^{n-1}(1),
G_3 = S_N \bigotimes C_n \bigotimes U_3(1)$ where
$C_r = \{c_{kr}  = exp(i2\pi k/r): a^+_j \rightarrow c_{kr} a^+_j, k = 0,1...,
r-1\}
, S_N$ is
the permutation group, and continious subgroups $U_j(1) = exp(i\lambda_j R_j)$
chara
cterize
symmetries of the interaction of two subsystems and are generated by the
integrals o
f motion $R_j:
R_1  = (m a^+_1 a_1 + n a^+_2 a_2)/(m+n),  R_2^0  = (m \sum_{j=1}^n a^+_j a_j +
n a^
+_0 a_0)/(1+n),
R_2^{\alpha}= a_{\alpha}^{+}a_{\alpha} - a_{\alpha +1}^{+}a_{\alpha +1}, \alpha
=1,.
..,n-1,
 R_3 = \sum _{i=1}^N \sigma_0 (i) + a^+_1 a_1/n $. This occurence of the
Hamiltonian
 symmetries
 enables us to introduce  in models (2.1) $G_i$-invariant collective  dynamical
vari
ables
 $Y_{\alpha}, \alpha = 0,+,-, Y_- = (Y_+)^+$ via a generalized Jordan-Schwinger
 mapping[8,9], namely, $ Y_0  = (a^+_1 a_1 - a^+_2 a_2)/(m+n),  Y_+ = (a_1^+
)^n(a_2
)^m$ for $H_1$,
 $ Y_{0} = (\sum_{i=1}^{n}a^{+}_{i} a_{i} - a_{0}^{+}a_{0})/(m+n),
Y_{+} = a_{1}^{+}...a_{n}^{+}(a_{0})^{m}$  for $H_2$
  and $Y_{0} = \sum_{i=1}^{N}\sigma_{0}(i),
Y_{+} = \sum_{i=1}^{N} \sigma_{+}(i)(a_{1})^{n}$  for$ H_3$. (We note that, in
gener
al,  a choice of
$G_i$-invariant collective  dynamical variables  $Y_{\alpha}$ is not unique and
can
be modified
(from physical considerations) by adding arbitrary functions of integrals of
motion
$R_j$ in their definitions.)
Then, one can rewrite all Hamiltonians  $H_i$  from (2.1) in the form
%============================================================================
$$    H  =  a Y_{0}  +  g Y_{+}  + g^{*} Y_{-}  + C ,
{}~~~~~~~~~      [Y_{\alpha } , C ] = 0     \eqno   (2.2) $$
%============================================================================
where the constant $a$ is determined by frequency detuning and $C$ is an
integral
of motion depending on the above $G_i$ generators $R_j$; for example, for $H_1$
at $
m=1, n=2$ we
have $a = 2\omega_1 - \omega_2, C = R_1(\omega_1 + \omega_2)$.
Operators $Y_{\alpha }$  are generators of Lie-like algebras
$g_{pd} (H)$ and satisfy CRs
%===========================================================================
$$  [Y_{0} , Y_{\pm }] = \pm Y_{\pm },~~~~
[Y_{-} , Y_{+} ] = \psi(Y_{0} ) =  \phi(Y_{0} +1) -
\phi(Y_{0})  \eqno (2.3)  $$
%============================================================================
where $ \phi(Y_{0})$ are  appropriate structure polynomials of the algebras
$g_{pd} (H)$ determined from the above definitions of $Y_{0}$ and
properties of operators $a^+_i, a_j, \sigma_{\alpha}(i)$; specifically, we have
%========
$$\phi(Y_{0})=\phi_1(Y_{0} ) =(2 Y_{0} + R_1)(2 Y_{0} + R_1-1) (R_1 - Y_{0} +
1) \eq
no (2.4a)$$
for $H_1$ at $m=1, n=2$,
 $$\phi(Y_{0})= \phi_2(Y_{0}) = [R_2^0 - Y_{0} + 1]([R_2^{0} + R_2^{1}]/2 +
Y_{0})
 ([R_2^{0} - R_2^{1}]/2 + Y_{0})
\eqno (2.4b)$$
for $H_2$ at $m=1, n=2$,
$$\phi(Y_{0})= \phi_3(Y_{0}) = [C_{2}(2) - Y_{0}^{(2)}][nR_3-nY_{0} + n]^{(n)},
A^{(
B)} =A(A-1)...(A-B+1)
\eqno (2.4c)$$
%==========
for $H_3$ where $C_{2}(2)=\Sigma_+ \Sigma_- + \Sigma_0^{(2)}$ is the Casimir
operato
r of the usual Lie algebra $su(2)$
with generators $\Sigma_{\alpha}=\sum_{i}\sigma_{\alpha}(i)$ describing the
atom
subsystem, and $[C_{2}(2), Y_{\alpha}] = 0$ . We note that variables $ R_j$ and
$Y_0$ determine populations of single field modes and of the atom subsystem
whereas
$Y_{\pm}$ are transition operators between states with fixed energies of
uncoupled
subsystems.

Since relations (2.3) resemble the CR for the familiar Lie algebra $sl(2)$ one
may
identify $g_{pd}(H)$ as different (distinguished by structure polynomials)
deformati
ons
$sl_{pd}^{\phi_{\alpha}}(2)$ of the Lie algebra $sl(2)$; herwith for $H_1$ and
$H_2$
 algebras
$sl_{pd}^{\phi_{\alpha}}(2)$
have two (noncompact and compact) versions distinguished by some features of
their irreps on $L(H)$ and
referred to as $su_{pd}^{(n)} (1,1)$ (for $m=0$) and $su_{pd}^{(n,m)} (2)$ (for
$m \neq 0$) whereas for $H_3$ there are only compact versions
$su_{pd}^{\phi_{3}}(2)
$.
    It is easy to check that all these algebras $sl_{pd}^{\phi_{\alpha}} (2)$
have t
he
Casimir operators
%==
$$C_{2}^{\phi} (2) = - Y_{+} Y_{-}  +  \phi (Y_{0} ) \eqno (2.5)$$
%===
that is a specific deformation of the usual $sl(2)$ Casimir operator
$C_{2} (2)= \pm E_{+} E_{-} + E_{0}^{(2)}$ ($E_{i}$ are the $sl(2)$ generators)
 [1]
. This allows us to develop a theory of the
$sl_{pd}^{\phi} (2)$ representations by analogy with that of usual Lie algebras
[1,2
].
Furthermore,  because of $C_{2}^{\phi} (2) = 0$ on  $L(H)$[8,9],
the algebras $sl_{pd}^{\phi_{\alpha}}(2)$ together with groups $G_i$
form generalized Weyl-Howe dual pairs on the spaces $L(H)$ that leads to the
decompo
sitions
%======
$$ L(H)  = \sum_{[l_i]} \oplus L([l_i ]) \eqno (2.6)$$
%====
of $L(H)$ into direct sums of the subspaces $L([l_i ])$ which are
invariant with respect to actions of both $G_i$  and $sl_{pd}^{\phi _i} (2)$.
The
complex label $[l_i]=(l_0,l_1,...,l_p)$ specifies irreducible representations
(irreps) of both $G_i$  and $sl_{pd}^{\phi _i} (2)$
simultaneously; herewith $l_0$ is the $sl_{pd}^{\phi} (2)$ lowest weight, and
other
numbers $l_i, i=1,..., $ are integrals of motion related to eigenvalues of the
above
mentioned
$U_j(1)$ generators $R_j$. We note that from physical point of view the
decompositio
n
(2.6) corresponds to a splitting of complex quantum systems into a set of
specific
"domain" subsystems evolving independently, i.e., if a wave vector $|\phi(t)>$
of th
e system
belongs to the subspace $L([l_i])$ at an initial time $t_0$, then it will
belong
to the same space $L([l_i])$ at any other time $t$.

The subspaces $L([l_i ])$ are spanned by orthonormalized
 basis vectors $|[l_i];v>= N([l_i],v)(Y_+)^v |[l_i]>$
where $(N([l_i],v))^{-2}=<[l_i]|(Y_-)^v (Y_+)^v |[l_i]>=\prod _{r=0}^{v-1}
\phi(l_0+v-r) \equiv [\phi(l_0+v)]^{(v)}$ and $|[l_i]>$ are the lowest vectors
which are defined by the conditions
%=====
$$Y_{0} |[l_{i}]> = l_{0} |[l_{i}]>, Y_{-} |[l_{i} ]> = 0 \eqno (2.7)$$
%============
and are given for different $H_i$ as follows
%=======================
$$a) H_1, m=1, n=2: |[l_i]>=[k!s!]^{-1/2}(a_{1}^{+} )^{k} (a_{2}^{+})^{s}  |0>,
l_{0} = (k-s)/3, l_{1} = (k+2s)/3,$$
$$R_1 |[l_{i} ]> = l_{1} |[l_{i} ]>,
k= 0,1,  s= 0,1,...,    \eqno (2.8a)$$
$$b) H_2, m=1, n=2:  |[l_i]>=[k!s!]^{-1/2}(a_{1}^{+})^{(k+l_2)/2}
(a_{2}^{+})^{(k-l_
2)/2} (a_{0}^{+})^{s}|0>,
l_{0} = (k-s)/3,$$
$$l_{1} = (k+2s)/3, R_2^0 |[l_{i}]> = l_{1} |[l_{i} ]>,
 l_2 =\pm k,  R_2^{1} |[l_{i} ]> = l_{2} |[l_{i}]>,
k= 0,1,..., s= 0,1,...,  \eqno (2.8b)$$
$$c) H_3: |[l_i]>=[k!]^{-1/2}(a_{1}^{+} )^{k} |0> |-j; \{j_{int}\}>,
l_{0} = -j, l_{1} = k/n -j,$$
$$R_3 |[l_{i} ]> = l_{1} |[l_{i} ]>,
 k= 0,1,..., 0 \leq j \leq N/2,    \eqno (2.8c)$$
%====================================
where $|-j; \{j_{int}\}>$ are the lowest vectors of the "atom" group  $SU(2)^a$
irre
ps
$D^j$ which are obtained from $\prod _{i=1}^N |\pm>(i)$ with the help of the
general
ized
Wigner coefficients and a set $ \{j_{int}\}$ of the  $SU(2)^a$ intermediate
angular
momenta labels
basis vectors of the irreps of the symmetric group $S_N$ which are dual to the
irrep
s $D^j$ on  subspaces
$L([l_i ])$[9]. Note that from the physical viewpoint  this set $ \{j_{int}\}$
of ex
tra
integrals of motion of Hamiltonians $H_3$ as well as some specific quantum
numbers (
"parities")
related to discrete symmetries $C_p$ of the structure
polynomials $\phi(l_0 +v)$ in models (2.1b) (see (2.4b) and (2.8b))
 characterize a specific "hidden supersymmetry" (cf. [11]) since they are
absent
in the $G_i$-invariant expression (2.2) for $H_i$ and in eqs (2.4) for
$\phi(l_0 +v)
$, and,
therefore,different subspaces $L([l_i ])$ distinguished
by these quantum numbers only have (up to constants depending on $l_i$) the
same ene
rgy spectra[9].

Actions of operators $Y_{\alpha}$ on the vectors $|[l_i];v>$ are given
as follows
%===========
$$ Y_0 |[l_i];v> =(l_0 + v)|[l_i];v>,~~~ Y_+ |[l_i];v> = [\phi(l_0 +v+1)]^{1/2}
|[l_
i];v+1>,$$
$$Y_- |[l_i];v> = [\phi(l_0 +v)]^{1/2} |[l_i];v-1>, |[l_i];0>\equiv |[l_i]>
\eqno (
2.9)$$
%==========================
Evidently, the algebras $su_{pd}^{(n)} (1,1)$ have on the space
$ L(H) = L_{F} (1)$ only "n" infinite-dimensional irreps $D([l_{0}])$
whereas the algebras $su_{pd}^{\phi} (2)$ have on $L({H})$
an infinite number of finite-dimensional irreps $D([l_{i}])$ which are
two-side bounded (with the highest vectors $|[l_i];M>, Y_0 |[l_i];M> =
(l_0 + M) |[l_i];M>, Y_+ |[l_i];M> =0$ and dimensions $d([l_i])= M+1$).
Besides,
from (2.8) and $C_{2}^{\phi} (2) = 0$ on  $L(H)$ one gets identities
%======
$$\phi(l_0 ) \equiv 0 \equiv \phi(l_0 +M+1) \eqno (2.10)$$
%========
In turn, these identities allow us to define a "radial" operator $E_r =
\sqrt{\phi(Y
_0)}$
by analogy with the case of the usual $sl(2)$ algebra[12] in a "polar"
decomposition

%===
$$a) Y_+ = E_r E, Y_-  = E^+ E_r,~~~~~~~~~~~  b) E^+ = E^{-1}   \eqno (2.11)$$
%===
of the  $sl_{pd}^{\phi} (2)$ shift operators $Y_{\pm}$ where a "phase" operator
$E$
can be determined (with taking into account (2.9) and (2.10)) by the difference
oper
ator $\Delta_{v}$
acting on the label variable $"v"$ of basis vectors $|[l_i];v>: E\equiv E_v =
\Delta
_{v} + 1, E |[l_i];v> =
|[l_i];v + 1>$. We note that the definition (2.11a) of the phase operators $E$,
in f
act,
does not determine their actions on the extremal vectors $|[l_i]> and
|[l_i];M>$ of
$L([l_{i}])$; therefore, it is necessary to add some conventions for "closing"
actio
ns of
the operators $E, E^+$ on $L([l_{i}])$. For example, by analogy with the usual
$sl(2
)$ algebra[12]
one may use standard "cyclic" conditions: $|[l_i]; M+1+v> = |[l_i]; v>$ for
$su_{pd}
^{\phi}(2)$
and  $|[l_i];-1> = |[l_i]; \infty>$ for $su_{pd}^{\phi}(1,1)$  which are in
accordan
ce
with (2.11b)). (A more detail discussion of this point is beyond the scope of
our pa
per.)

\section{Evolution operators for quantum systems with dynamic symmetry algebras
 $sl_{pd}(2)$}
Now we discuss applications of the  $sl_{pd}(2)$ algebras for solving
physical problems with Hamiltonians (2.2) (from hereon we omit the superscript
$"\ph
i"$ in
notations of polynomial Lie algebras unless when it is unnecessary). For lack
of
simple formulas for disentangling exponents $\exp (\sum d_{i}  Y_{i} )$ [8,13]
one cannot apply the orbit coherent states techniques (or simililar ones) [2]
for diagonalizing $H$  or for finding appropriate evolution
operators $U_{H} (t)$ as it is the case for usual Lie algebras.
Nevertheless, there exist some possibilities of applications
of the $sl_{pd}(2)$ formalism to solving these tasks which are based
on using expansions of the energy eigenfunctions $|E>$, the evoluton operators
$U_{H}(t)$ (in the Schroedinger picture) or collective dynamical variables
$Y_{\alph
a}(t)
(Y_{\alpha}(0) = Y_{\alpha}\in sl_{pd}(2))$ (in the Heisenberg picture)
by power series in $Y_{\alpha}\in sl_{pd}(2))$  as well as on applying CRs
(2.3).

Particularly, for solving spectral problems this way led to a definition of new
classes of orthogonal polynomials (related to abelian functions)[9] as
solutions of

difference Schroedinger equations of the type which is widely used in studies
of
different nonlinear problems [14].
In the case of the temporal evolution problem analysis  within this approach
one can use  a representation [9]
%====
$$ U_H(t)\equiv  U_H(t;0) = \sum _{n\geq 0}[A_n (Y_0;t) (Y_-)^n  +  (Y_+)^n
B_n(Y_0;
t)], A_0 (Y_0;t) \equiv 0,
\eqno (3.1)$$
%====
for evolution operator $U_H(t)$ with the Hamiltonian (2.2) where for algebras
$su_{p
d}(2)$
series in (3.1) are terminating. Then, substituting the ansatz (3.1) and Eqs
(2.2)
 in the Schroedinger equation $i\hbar \partial_{t} U_H(t) = H U_H(t)$ for
 $U_H(t)$ ($\partial_{t} \equiv \partial/ \partial t$) and using Eq. (2.5)
together
with the condition,
$C_{2}^{\phi} (2) = 0$ on  $L(H)$,
 one gets after some algebra a system of differential-difference equations
 %===
 $$a)i\hbar\partial_{t} A_k(Y_0;t) =(aY_0 + C) A_k(Y_0;t) +$$
 $$g  A_{k+1}(Y_0-1;t)\phi (Y_0) +
 g^* A_{k-1}(Y_0+1;t) + g^*  B_0(Y_0+1;t) \delta_{k,1} , k=1,2,...,      \eqno
(3.2a
)$$
 $$b)i\hbar\partial_{t} B_k(Y_0;t) =(aY_0 +ak+ C) B_k(Y_0;t) +$$
 $$g  B_{k-1}(Y_0;t)+
 g^* B_{k+1}(Y_0;t)\phi (Y_0+1+k)  , k=0,1,2,...      \eqno         (3.2b)$$
%====
which can be can be interpreted as Schroedinger equations with difference
Schroeding
er
operators in variables $"k"$ and $Y_0$. These equations should be solved
together wi
th
initial conditions $A_k(Y_0;t=0) = 0, B_k(Y_0;t=0) = \delta _{k,0}$ following
from $U_H(t=0)= I$ (a Cauchy problem) on each subspace $L([l_i ])$ separately.
Below we develop a general scheme of solving this problem  exploiting  some of
its
features.

First, we note that for  determining the total operator $U_H(t)$ it is
sufficiently,
 in principle,
to solve equations (3.2b) for the $c$-number functions $B_{k}(l_0; t) $
obtained by acting $U_H(t)$ on the lowest vector $|[l_{i} ]>$ as an initial
state $|
\psi (t=0)>$.
Indeed, using the semigroup property of $U_H(t)$ and knowing the functions
$B_{k}(l_
0; t) $
one can determine the action of $U_H(t)$ on the "generalized coherent" states
$|\psi(\tau)> = U_H(\tau)|[l_{i}]> = \sum _{n\geq 0} (Y_+)^n
B_n(l_0;\tau)|[l_{i}> $

from the equation
%===
$$ U_H(t) |\psi(\tau)> = U_H(t)U_H(\tau)|[l_{i}]> = U_H(\tau + t)|[l_{i}]>
\eqno (3.
3)$$
Performing multiple differentiations of (3.3) with respect to $\tau$ and then
putting $\tau =0$ one gets a set of equations
%======
$$  \sum_{r\geq 0}\gamma_{rp} U_H(t) (Y_+)^r|[l_{i}]> =
\sum_{r\geq 0}(\partial_t)^p B_{r}(l_0; t)(Y_+)^r|[l_{i}]>,
\gamma_{rp}= (\partial_t)^p B_{r}(l_0; t)|_{t=0}, p=0,1,... \eqno (3.4a)$$
%====
defining actions of $U_H(t)$ on any basic vectors $|[l_{i}]; v>$:
%===
$$  U_H(t) (Y_+)^v|[l_{i}]> = \sum_{r,p} \gamma_{pv}^{-1}(\partial_t)^p
B_{r}(l_0; t
)(Y_+)^r|[l_{i}]>
\eqno (3.4b)$$
%========
Note that the matrix
$\Gamma = ||\gamma_{rp}||$ is the Gaussian one  as it follows from (3.2b) and
the initial condition $B_k(Y_0;t=0) = \delta _{k,0}$, and, therefore, is easily
inverted. For example, $U_H(t) Y_+ |[l_{i}]>=ig^{-1} \partial_{t} |\psi(t)>;
U_H(t) (Y_+)^2 |[l_{i}]>=-(g^{-2} \partial_{t}^2 + g^*g^{-1} [\phi(l_0+2) +
\phi(l_0
+1)])
|\psi(t)>$.

Second, when structure polynomials $\phi (x)$ have degree $deg \phi \leq 2$,
 solutions of eqs. (3.2) can be  easily obtained from  known expressions for
$U_H(t)$[1,2,15] since in these cases  algebras  $sl_{pd}^{\phi} (2)$ are
reduced
to usual Lie algebras. Therefore, one may develop calculating schemes  using
such
"quadratic" solutions as some auxiliary "initial" ones for the original
problem.
Specifically, in [8,9], we outlined  iterative
schemes of finding  functions $B_{k}(l_0; t) $ using as a zero-th order
approximatio
ns
well-known[1,2,15] solutions of eqs (3.2b) when the original structure
polynomial $
\phi (x)$
being replaced by  a certain quadratic polynomial $\phi_2 (x)$.
These approximations correspond to replacing  the original  Hamiltonians (2.2)
by
"distorted" Hamiltonians $H_{D} =\bar{a} V_{0} + \bar{g} V_{+} + \bar{g^{*}}
V_{-} +
 \bar{C},
[\bar{C}, V_{\alpha}] =0$, where $V_{\alpha}$ are generators of usual
three-dimensio
nal
Lie algebras $g^D$ connected with $sl_{pd}(2)$ via a generalized
Holstein-Primakoff
mapping, and coefficients $\bar{a}, \bar{g}$ are  chosen from conditions of
maximal proximity Hamiltonians $H$ and $H_D$. But in this paper we will exploit
anot
her
idea related to a specific recursive mapping of solutions of certain auxiliary
"quad
ratic"
task onto solutions of the Cauchy problem (3.2b) for $B_{k}(l_0; t) $.
Herewith, in

order to elucidate the essence of our approach we consider in detail the case
when
 $deg \phi=3 $ and restrict ourselves only by analysis of the case of the
compact
 algebra $su_{pd}(2)$ (that corresponds to $n=2, m=1$ in (2.1a), (2.1b) and
$n=1$ in
 (2.1c)),
 since in this case our approach is most efficient.

 So, we look for solutions of the following Cauchy problem
 %==
 $$i\dot B_k(l_0;t)-
 [(al_0 +ak+ c) B_k(l_0;t) + g  B_{k-1}(l_0;t)+
 g B_{k+1}(l_0;t)\phi (l_0+1+k)]=$$
 $$\delta(t) \delta_{k,0},~~~ k=0,1,2,...,   \eqno         (3.5)$$
%====
where $\dot B_k(l_0;t) \equiv \partial_{t} B_k(l_0;t), c$ is the eigenvalue
of the operator $C$ in (2.2) on $L([l_i])$, the initial condition
$ B_k(l_0;t=0) = \delta _{k,0}$ is accumulated in rhs of (3.5) and for the
sake of simplicity we put $\hbar=1$.
Represent the polynom $\phi(x)$ in the form
%==
$$\phi(x)=\phi_2(x)(\alpha+\gamma x),~ ~  \phi_2(l_0)=0=\phi_2(l_0+M+1)
\eqno(3.6)$$

%=====
and use in in (3.5)  the substitution
%==
$$B_k(l_0;t)=\frac{b_k(t)}{(\tilde{\alpha} g^*)^k [\phi_2(l_0+k)]^{(k)}},~~
k=0,1,2,
...,
$$
$$
[\phi_2(l_0+k)]^{(k)} =\prod _{r=0}^{k-1} \phi_2(l_0+k-r)],
\phi_2(l_0+k)]^{(0)} =1

\eqno (3.7)$$
%==
where $\tilde{\alpha}=\alpha$ for $\alpha \neq 0$ and $\tilde{\alpha}=1$ for
$\alpha =0$ (that is the case, e.g., for the model (2.1b) on the subspaces
$L(l_0=-s/3,l_1=2s/3,l_2=0)$ with a typical physical initial state
$|\phi(t=0)>=
|0,0,\nu_0> =exp[\nu_0 a^+_0 - \nu_0^*a_0]|0>$). Then one finds
%==
 $$i\dot b_k(t)-
 [(al_0 +ak+ c)b_k(t) + |g|^2\alpha b_{k-1}(t) \phi_2(l_0+k)+
  b_{k+1}(t)(1+\beta (k+1))]=
 \delta(t) \delta_{k,0}, $$  $$k=0,1,2,...,~~ \beta =\gamma/\alpha
 \eqno   (3.8)$$
 %===
for $\alpha \neq 0$ and a similar (with replacing "$(1+\beta (k+1))$" by
"$\gamma (k+1)$") equation for $\alpha =0$.
Now, implementing in (3.8) the temporal Fourier transform
%===
$$ b_{k}(t) =  \frac{1}{2\pi} \int_{-\infty}^{+\infty} exp(-i\omega t)
\hat{b}_{k} (\omega) d \omega \eqno (3.9)$$
%===========
and introducing the notation
%======
$$ \hat{L} =\omega - [(al_0 +ak+ c)+ |g|^2 \alpha \phi_2(l_0+k) E_k^{-1}+
 g E_k],~~~ E_k= \Delta_{k} + 1 \eqno (3.10)$$
 %===========
one gets a finite-difference equation
%===
$$ \hat{L}\hat{b}_{k} (\omega) = \delta_{k,0}+ \beta (k+1) E_k
\hat{b}_{k}(\omega)
\eqno (3.11)$$
%===
whose solutions yield those (for $\alpha \neq 0$) of the original task (3.5)
and, consequently, of (3.2b).

Suppose now that we know solutions of the task
%===
$$ \hat{L}\hat{b}_{k}^0 (\omega) = \delta_{k,0},~~~ \hat{b}_{-1}^0 (\omega) =
0=\hat
{b}_{M+1}^0 (\omega)
\eqno (3.12a)$$
%============
Then, as is well-known from the functional analysis,  for finding solutions of
Eqs (
3.11) there is a standard
iterative calculation scheme
%+++=======
$$\hat{b}_{k} (\omega) = \hat{b}_{k}^0 (\omega) + \sum_{n\geq0} \hat{b}_{k-n}^0
(\om
ega)
\beta (n+1) \hat{b}_{n+1} (\omega) \eqno (3.13)$$
%===========
Now we use the so-called method of the $z$-transform[16] which is an effective
 tool for solving finite-difference equations. Specifically, applying to (3.13)
the
$z$-transform
%===================
$$\hat{b}_{k} (\omega) \longmapsto \hat{B}(z;\omega)=\sum_{k=0}^\infty
\hat{b}_{k}(\
omega) z^{-k}
\eqno (3.14)$$
%===========
and using the standard techniques of the method of the $z$-transform[16] one
gets
a first-order differential equation
%=========
$$\hat{B}(z;\omega) = \hat{B}^0(z;\omega)(1- \beta z^2 d/dz \hat{B}(z;\omega))
\eqno (3.15)$$
%===
where $\hat{B}^0(z;\omega)$ is the $z$-image of the solution of Eqs (3.12a). A
gener
al
solution of Eq. (3.15) has the form
%=====++++=====
$$\hat{B}(z;\omega) = \Phi(z^{-1};\omega)
 [-(\beta)^{-1} \int_{0}^{z^{-1}}\frac{dx}{ \Phi(x;\omega)}+ D(\omega)]
 \eqno (3.16a)$$
%===========
where
%===
$$ \Phi(z;\omega)= exp[(\beta)^{-1}
\int_{0}^{z}\frac{dx}{\hat{B}^0(x^{-1};\omega)}]

\eqno (3.16b)$$
%=====
and the integration constant in (3.16) $D=\hat{B}(z=\infty;\omega)$ coincides
with
$\hat{b}_0(\omega)$. All other quantities $\hat{b}_{k}(\omega)$ are expresed in
terms of $\hat{b}_0(\omega)$ and can be found from (3.16) with the help of
inversion

of the map (3.14) using, e. g., the Taylor formula[16]
%=============
$$\hat{b}_{k} (\omega) = \frac{1}{k!}[\frac{d^k
\hat{B}(z^{-1};\omega)}{dz^k}]_{z=0}

(k=1,...)     \eqno (3.17)$$
%=====================
In turn, $\hat{b}_0(\omega)$ can be determined from the additional condition
$\hat{b}_{k=M+1}(\omega)=0$ which is consistent with the terminating character
of the series (3.1) for $su_{pd}(2)$. For example, in the case of 2-dimensional
subspaces $L([l_i])$, when $M=1$, we find $\hat{b}_1(\omega)=
(\hat{b}_0(\omega)/\hat{b}_0^0(\omega)-1)\beta^{-1},
\hat{b}_0(\omega)=\hat{b}_0^0(\omega)/(1- \beta \hat{b}_1^0(\omega))$. (Note
that
one can develop a procedure like Eqs (3.11)-(3.17) without using the Fourier
transfo
rm
(3.9); however, on this way we get an integral - differential equation for
$\hat{B}(z;l_0,t)= \sum_{k=0}^\infty \hat{B}_{k}(l_0;t) z^k$ instead of the
fairly
simple Eq.(3.15).)

Thus, for $\alpha \neq 0$ we have reduced the original task (3.2b) to solving
Eqs (3
.12a) that, in turn, is determined
(according to general theory of operators)  by solutions of the Sturm-Liouville
prob
lem
%===
$$(\omega - al_0 - c - \hat{L})\hat{P}_{k}(\Omega) = \Omega
\hat{P}_{k}(\Omega), ~~~
\hat{P}_{-1}(\Omega) = 0=\hat{P}_{M+1} (\Omega)
\eqno (3.12b)$$
%============
with the finite-difference Liouville operator (3.10):
%=========
$$\hat{b}_{k}^0 (\omega) =N_{0} \sum_{r=0}^M\frac{\hat{P}_{k}(\Omega_r)
\hat{P}_{0}(
\Omega_r)}
{\omega - \Omega_r - al_0 - c},~~~ N_{j} \sum_{r=0}^M\hat{P}_{k}(\Omega_r)
\hat{P}_{
j}(\Omega_r) =
\delta_{kj} \eqno (3.18)$$
%==========
where $\Omega_r$ and $\hat{P}_{k}(\Omega_r)$ are eigenfrequencies and
eigensolutions
 of
the problem (3.12b), respectively. As  in (3.10) $deg \phi=2 $, the solutions
of this problem for $\alpha > 0$ are well-known[9,15,17] and are related to
diagonal
ization of
an arbitrary element of the usual $su(2)$
algebra. Specifically, its frequency spectrum is linear:
$\Omega_r = \Omega_r^0 + \Omega^{'} r$ ($\Omega^{'}$ is a constant),
$\hat{P}_{k}(\O
mega_r)$ are classical orthogonal
polynomials in the discrete variable $\Omega_r$, related to values of the
Gaussian
hypergeometric function $_2 F_1(a,b; c; z)$ or the $su(2)$ $d$-function at
certain f
ixed values
of their continious argument $"z"$, and $\hat{B}^0(z^{-1};\omega)$ are
expressed in
terms
of the standard generating functions for these polynomials[15,17].
In the case of $\alpha < 0$ the procedure above is slightly modified and is
related
to
the diagonalization of an arbitrary element of $su(1,1)$ on non-unitary
finite-dimen
sional
representations that leads to purely imaginary eigenfrequencies $\Omega_r$.
For example, in the case of the model (2.1a) with $m=1, n=2, \omega_2
=2\omega_1$
and $k=0$ in Eqs (2.8a), we have $\phi (x) = 2x(s+1-x)(2x-1)$ in (3.6) and
$\alpha=-
2, \beta =-2$.
Herewith Eq. (3.18) is specified (for $g=g^*$) as follows
%===
$$\hat{b}_{k}^0 (\omega)/(g)^k =\sqrt{s^{(k)} k!2^k}
\sum_{r=0}^s\frac{P_{r-s/2,k-s/
2}^{s/2}(0) P_{r-s/2,-s/2}^{s/2*}(0)}
{\omega - \Omega_r -2s\omega_1}, ~~~ \sum_{r=0}^sP^{s/2}_{r-s/2,k}(z)
P^{s/2*}_{r-s/
2,j}(z) =
\delta_ {kj} \eqno (3.19)$$
%===
where  $\Omega_r = i2\sqrt{2}|g|(r-s/2)$  and $P_{r-s/2,k-s/2}^{s/2}(0) $ are
the $s
u(2)$ $d$-functions at
$z=0$[17].

 So, we have obtained an procedure to solve the problem (3.5) for $su_{pd}(2)$
in th
e
case when $deg \phi=3 $ and $\alpha \neq 0$ in (3.7). Evidently, considering
its sol
utions to be found one may
obtain those for the case $deg \phi=4 $ and, generally speaking, generalize
the algorithm above for $deg \phi=n $ where $n$ is an arbitrary natural number.
For $\alpha = 0$ the procedure is similar though the analogs of Eqs (3.15) and
(3.16
)
become in somewhat more complicated.
As for the  case of noncompact  algebras $su_{pd}(1,1)$ (models (2.1a) and
(2.1b) at
 $m=0$)
we note that this approach is  less efficient since a problem of determining
the "i
ntegration
constant" $D(\omega)= \hat{b}_0(\omega)$ in (3.16) is open: due to the absence
of an
y terminating condition for
$\hat{b}_{k}(\omega)$ we can use only integral spectral conditions
%=====
$$\int_{-\infty}^{+\infty}
\hat{b}_{k} (\omega) d \omega =2\pi\delta_{k0}, k=0,1,2,...  \eqno (3.20)$$
 %===============
following from (3.9) and $B_{k}(l_0;0)= \delta_{k0}$. As a whole, a further
advance
in developing the approach proposed
for both compact and non-compact version of $sl_{pd}(2)$ appears to be
connected wit
h
a thorough analysis of analitycal properties of the "spectral
function"$\Phi(z^{-1};
\omega)$.

\section{Conclusion}

  In conclusion we point out that a similar approach can be used and for
solving
  the spectral problem of the (3.12b) type ( but with $deg \phi\geq 3$) for
finding
  the energy spectra  and appropriate eigenfunctions (cf. [8,9]). Specifically,
 on this way we get in the case, when $\alpha\neq 0$, a generating function
$\hat{Q}
(z;\lambda)$
 for the coefficients $Q_{k}(\lambda(E))$ of the expansions $|E> =\sum_{f\geq
0}
 Q_{f}(\lambda(E))(Y_+)^f|[l_i]>$ in the form
 %======
 $$\hat{Q}(z;\lambda) =\Phi(z^{-1};\lambda)   \eqno (4.1)$$
 %========
 Herewith in the case of the compact algebra $su_{pd}(2)$,
when the constants $Q_{0}(\lambda(E))$ are found from the $|E>$ normalization
conditions[8,9], energy spectra are determined from the condition
 %=====
 $$[\frac{d^{M+1} \hat{Q}(z^{-1};\lambda)}{dz^{M+1}}]_{z=0} =0 \eqno(4.2a)$$
 %=========
 or
 %===
 $$ \oint dz z^M \hat{Q} (z;\lambda) =0    \eqno (4.2b)$$
 %====
 where the integration contour in (4.2b) encloses all the $\hat{Q} (z;\lambda)$
sing
ularities[16].
 For $M\gg 1$ the form (4.2b) allows us to facilitate the determination of
spectra w
ith
 the help of the saddle point method.
 Note that a closed form (4.1), (3.16b) for functions $\hat{Q}(z;\lambda)$ has
some
 advantages in comparison with their implicit  determinations in [8,9] in order
to e
xamine
 some analytical properties of $Q_{k}(\lambda(E))$  and $\{E_l\}$.
 It is also of interest to generalize the analysis above for
 multimode versions of the Hamiltonians (2.1) related to
polynomial Lie algebras $g_{pd}$  with the coset structure $g_{pd}= h +v$ and
$h = u(m)$ [8,9]. But for this end we need an additional work for separating
variables. Besides, using generalized Holstein-Primakoff mappings [9]
one can determine some relations between familiar Lie algebras and both
polynmially and q-deformed Lie algebras $g_{d}$ in order to display
different exotic states and phenomena[18] in realistic multi-particle models
as well as to determine their asymptotic behaviours [9].
\section {Acknowledgements}
The authors thank S.M. Chumakov for stimulating discussions. V.P.K. is
 grateful to the staff of Facultad de Ciencias Fisico-Matematicas de la
Universidad
 de Guadalajara for its hospitality during preparing this work,
 and A.K. acknowledges support from CONACYT of Mexico.

\end{document}